\documentclass[runningheads,fleqn]{cl2emult}

\usepackage{makeidx}  
\usepackage{graphicx} 
\usepackage{subeqnar} 
\usepackage{multicol} 
\usepackage{cropmark} 
\usepackage{comp}     
\makeindex            

\usepackage{alltt}
\usepackage{moreverb}
\usepackage{latexsym}
\usepackage{amsfonts}

\newcommand{\eventset}{\mathbb{E}}
\newcommand{\traceset}{\mathbb{T}}
\newcommand{\code}[1]{{\tt #1}}

\newcommand{\codec}[1]{`{\tt #1}'}

\newcommand{\myemph}[1]{\code{#1}}

\newcommand{\fold}{\code{fold}}

\newcommand{\collect}{\code{foldt}}
\newcommand{\foldt}{\code{foldt}}

\newcommand{\initialize}{\code{initialize}}
\newcommand{\filter}{\code{collect}}
\newcommand{\postpro}{\code{post\_process}}

\begin{document}
\title*{Generic Program Monitoring\protect\newline 
by Trace Analysis\thanks{This work has been partially sponsored by 
the Esprit project ARGO, Industrial RTD Project no 25503}}
\toctitle{Generic Program Monitoring by Trace Analysis}
%
%
\titlerunning{Generic Program Monitoring by Trace Analysis}
%
\author{Erwan Jahier\inst{1}
\and Mireille Ducass\'{e}\inst{2}}
\authorrunning{Erwan Jahier and Mireille Ducass\'{e}}
%
%
\institute{ V\'{e}rimag, Centre Equation - 2 avenue de Vignate 38610 Gi\`{e}res
\and IRISA/INSA, Campus Universitaire de Beaulieu, F-35042 Rennes cedex, France}

\maketitle              

\begin{abstract}

\index{abstract}Program execution monitoring consists of checking
whole executions for given properties, and collecting global run-time
information.
Monitoring helps programmers maintain their programs.  However,
application developers face the following dilemma: either they use
existing monitoring tools which never exactly fit their needs, or
they invest a lot of effort to implement relevant monitoring code.
In this article we argue that, when an event-oriented tracer exists,
the compiler developers can enable the application developers to
easily code their own monitors.
We propose a high-level primitive, called \foldt, which operates on
execution traces.  One of the key advantages of our approach is that
it allows a clean separation of concerns; the definition of monitors
is totally distinct from both the user source code and the language
compiler.
We give a number of applications of the use of \foldt\ to define
monitors for Mercury program executions: execution profiles,
graphical abstract views, and test coverage measurements. Each
example is implemented by a few lines of Mercury.

\end{abstract}

\noindent
{\bf keywords:} monitoring, automated debugging, trace analysis, test coverage,
  Mercury.


\section{Introduction}

\paragraph{\bf Program maintenance and trace analysis.}

Several experimental studies (e.g., \cite{hatton97}) show that
maintenance is the most expensive phase of software development: the
initial development represents only 20~\% of the cost, whereas error
fixing and addition of new features after the first release
represent, each, 40~\% of the cost. Thus, 80~\% of the cost is due to
the maintenance phase.

A key issue of maintenance is program understanding. In order to fix
logical errors, programmers have to analyze their program symptoms and
understand how these symptoms have been produced. In order to fix
performance errors, programmers have to understand where the time is
spent in the programs.
In order to add new functionality, programmers have to understand how
the new parts will interact with the existing ones.

Program analysis tools help programmers understand programs. For
example, type checkers \cite{pfenning92} help understand data
inconsistencies. Slicing tools \cite{gallagher91,tip95} help
understand dependencies among parts of a program. Tracers give
insights into program executions \cite{eisenstadt88}.  

Some program analysis tools automatically analyze program execution
traces.  They can give very precise insights of program (mis)behavior.
We have shown how such trace analyzers can help users debug their
programs. In our automated debuggers, a  trace query
mechanism helps users check properties of parts of traced executions
in order to understand misbehavior
\cite{ducasse99c,ducasse99,morphine-ref}.

In this article, we show that trace analysis can be pushed toward
monitoring to further help understand program behavior.  
Whereas debuggers are tools that retrieve run-time information at
specific program points, monitors collect information relative to the
whole program executions.
For example, some monitors gather statistics which help detect
heavily used parts that need to be optimized; other monitors
build graphs (e.g., control flow graphs, dynamic call graphs, proof
trees) that give a global understanding of the execution.

\paragraph{\bf Execution monitoring.}

Monitors are trace analyzers which differ from debuggers. Monitoring
is mostly a ``batch'' activity whereas debugging is mostly an
interactive activity. In monitoring, a {\em set} of properties is
specified beforehand; the whole execution is checked; and the global
collected information is displayed. In debugging, the end-user is
central to the process; he specifies on the fly the very next
property to be checked; each query is induced by the user's current
understanding of the situation at the very moment it is typed in.
Monitoring is therefore less versatile than debugging.  The
properties specified for monitoring have a much longer lifetime,
they are meant to be used over several executions.

It is, nevertheless, impossible to foresee all the properties that
programmers may want to check on executions. One intrinsic reason is
that these properties are often driven by the application
domain. Therefore monitoring systems must provide some genericity.

\paragraph{\bf Existing approaches to implement monitors.}

Unfortunately, monitors are generally implemented by ad hoc
instrumentation. This instrumentation requires a significant
programming effort. When done at a low level, for example by
modifying the compiler and the runtime system, it requires deep
knowledge that mostly only the language compiler implementors
have. However, the monitored information is often
application-dependent, and application programmers or end-users know
better what has to be monitored. But instrumenting compilers is
almost impossible for them.

An alternative to low-level instrumentation is source-level
instrumentation; run-time behavior information can be extracted by
source-to-source transformation, as done for ML
\cite{tolmach95,kishon91} and Prolog \cite{ducasse2000} for instance
. Such instrumentation, although simpler than low-level compiler
instrumentation, can still be too complex for most programmers.
Furthermore, for certain new declarative programming languages like
Mercury~\cite{somogyi96}, instrumentation may even be impossible.
Indeed, in Mercury, the declarative semantics is simple and well
defined, but the operational semantics is complex. For example, the
compiler reorders goals according to its needs. Furthermore, input
and output can be made only in deterministic predicates. This
complicates code instrumentation.

Thus, ad hoc instrumentation is tedious at a low level and it may be
impossible at a high level.  On the other hand, the difficult task of
instrumenting the code to extract run-time information has, in
general, already been achieved to provide a debugger. Debuggers, which
help users locate faults in programs are based on tracers. These
tracers generate execution traces which provide a precise and faithful
image of the operational semantics of the traced language. These
traces often contain sufficient information to base monitors upon
them.

\paragraph{\bf Our Proposal.} 

In this article, we propose a high-level primitive built on top of an
event oriented execution tracer. The proposed monitoring primitive,
called \foldt, is a \fold\ which operates on a list of events.

An event oriented \emph{trace} is a sequence of events. An
\emph{event} is a tuple of event attributes.  An \emph{event
  attribute} is an elementary piece of information that can be
extracted from the current state of the program execution.  Thus, a trace
can be seen as a sequence of tuples of a database ordered by time.
Many tracers are event-oriented: for example, Prolog tracers based on
Byrd box model~\cite{byrd80}, tracers for C such as
Dalek~\cite{olsson90} and Coca~\cite{ducasse99c}, the Egadt tracer for
Pascal~\cite{fritzson94}, the Esa tracer for Ada~\cite{howden96}, and
the Ebba tracer for distributed systems~\cite{bates95}.

One of the key advantages of our approach is that it allows a clean
separation of concerns; the definition of the monitors is totally
distinct from both the user source code and the language compiler.

We have implemented  \foldt\ on top of the Mercury trace.
We give a number of applications of the \foldt\ operator to compute
various monitors: execution profiles, graphical abstract views, and 
test coverage measurements.
Each example is implemented by a few lines of Mercury which can be
written by any Mercury programmer.
These applications show that the Mercury trace, indeed, contains
enough information to build a wide variety of interesting monitors.
Detailed measurements show that, under some requirements, \foldt\ can
have acceptable performance for executions of several millions of
execution events.
Therefore our operator lays the foundation for a generic and
powerful monitoring environment.
The proposed scheme has been integrated into the Mercury environment.
It is fully operational and part of the Mercury distribution.

Note that we have implemented the \foldt\ operator on top of Mercury
mostly for historical reasons. We acknowledge that some of the
monitors were particularly easy to write thanks to the 
neatness
of Mercury libraries, in particular the set library (e.g.,
Figure~\ref{cfg-source}). Nevertheless, \foldt\ could be implemented
for any system with an event-oriented tracer.

\paragraph{\bf Plan.} 

In Section~\ref{collect-section}, we introduce the \foldt\ operator
and describe its current implementation on top of the Mercury tracer.  
In Section~\ref{collect-app}, we illustrate the genericity of \foldt\
with various kinds of monitors.  All the examples are presented at a
level of detail that does not presuppose any knowledge of Mercury.
Section~\ref{performance-section} discusses performance issues of
\foldt.
Section~\ref{collect-related} compares our contribution with related
work.  
A thorough description of the Mercury trace can be found in
Appendix~\ref{trace-appendix}.
Appendix~\ref{queens-prog-appendix} lists a Mercury program solving
the n~queens problem, which is used at various places in the
article as an input for our monitors.



\section{A high-level trace processing  operator: \foldt}
\label{collect-section}

In this section, we first define the \foldt\ operator over a general
trace in a language-independent manner. We describe an implementation
of this operator for Mercury program executions, and then present its
current user interface.

\subsection{Language independent  \foldt\ definition }
\label{formal-collect-section}

A trace is a list of events; analyzing a trace therefore requires to
process such a list.  The standard functional programming operator
\fold\ encapsulates a simple pattern of recursion for processing
lists. It takes as input arguments a function, a list, and an initial
value of an accumulator; it outputs the final value of the
accumulator; this final value is obtained by successively applying the
function to the current value of the accumulator and each element of
the list.
As demonstrated by Hutton \cite{hutton99}, \code{fold} has a great
expressive power for processing lists.  
Therefore, we propose a \fold-like operator to process execution
traces; we call this operator \foldt.

Before defining \foldt, we define the notions of event and trace
for sequential executions.

\begin{definition}(Execution event, Event attributes, Execution trace)
  \\ \emph{An execution event} is an element of the Cartesian product
  $\eventset = A_1~\times~...~\times~A_n $, where $A_i$ for $i \in
  \{1, ..., n\}$ are arbitrary sets called \emph{event attributes}.
  \emph{An execution trace} is a (finite or infinite) sequence of
  execution events; the set of all execution traces is denoted
  by~$\traceset$. 
  We note $|t|$ the size (its number of events) of a finite
  trace $t \in \traceset$ and $|t|=\infty$ the size of infinite
  traces.
\end{definition}

The following definition of \foldt\ is a predicative definition of a
\fold\ operating on a finite number of events of a (possibly infinite)
trace.
The set of predicates over $\tau_1~\times~...~\times~\tau_n$ is
denoted by $pred(\tau_1, ..., \tau_n)$.

\begin{definition}(\foldt)\\
\label{foldt-def}
  A  \emph{\foldt\ monitor of type $\tau \times
    \tau'$} is a 3-tuple : $(init, \filter, \postpro) 
  \in pred(\tau)~\times~pred(\eventset, \tau, \tau)~\times~pred(\tau, \tau') $ 
such that:
  $ \forall t=(e_i)_{i > 0} \in \traceset$, either \\
 
(1) 
$ |t|< \infty \ \land (\exists ! (V_0, ..., V_{n}) \in \tau^{n+1} .  $
\\
\phantom{xxxxxxxxx} $ (init(V_0) 
 \ \bigwedge_{i=1}^{n} \filter(e_i, V_{i-1}, V_{i}) \ \land\ 
 \postpro(V_{n}, Res)  ) )$ \\

(2)
$\exists ! n < |t|, \exists ! (V_0, ..., V_{n}) \in \tau^{n+1}, \forall x \in \tau
 . $
\\
\phantom{xxxxxxxxx} $ (init(V_0)  \ \land \ \bigwedge_{i=1}^{n} \filter(e_i, V_{i-1}, V_{i}) \ \land \ 
 \postpro(V_{n}, Res) \ $
\\
\phantom{xxxxxxxxx} $ \land \ \neg  \filter(e_{n+1}, V_n, x))$

\noindent
$Res$ is called \emph{the result of the monitor $(init,
  \filter, \postpro)$ on trace $t$}. We use the notation $\exists !
  n$ to mean that there exists a unique $n$, and $(e_i)_{i > 0}$ for
  the sequence (in the mathematical sense) $e_1, e_2, e_3, ...$.
\end{definition}

Operationally, an accumulator of type $\tau$ is used to gather the
collected information. It is first initialized ($V_0$).  The predicate
\filter\ is then applied to each event of the trace  in turn, updating
the accumulator along the way ($V_i$). 
There are two ways to stop this process: (1) the folding process stops
when the end of the execution is reached if the trace is finite ($|t|<
\infty$); (2) if \code{collect} fails before the end of the execution is
reached ($\forall x \in \tau . \ (\neg \filter(e_{n+1}, V_n, x))$).
In both cases,
the last value of the accumulator ($V_n$) is processed by \postpro,
which returns a value ($Res$) of type $\tau'$ ($\postpro(V_{n}, Res)$).

Note that this definition holds for finite and infinite traces (thanks
to the second case of Definition~\ref{foldt-def}). This is convenient
to analyze programs that run permanently. The ability to end the
\foldt\ process before the end of the execution is also convenient to
analyze executions part by part as explained in
Section~\ref{part-by-part-analysis-section}.
A further interesting property, which is useful to execute several monitors in
a single program execution, is the possibility to simultaneously apply
several \fold\ on the same list using a tuple of \fold~\cite{bird87};
in other words: \\
 
$ 
\foldt(i_1, c_1, p_1)~\times~...~\times~\foldt(i_n, c_n, p_n)
=\\ \phantom{xxxxxxxxxxxxxxxxxx}
\foldt(i_1~\times~...~\times~i_n, c_1~\times~... ~\times~c_n, p_1
\times ...~\times~p_n)$

where:

$ 
\forall a_1, ..., a_n \in \tau_1~\times~...~\times~\tau_n,
\\\phantom{xxxx}i_1~\times~...~\times~i_n(a_1, ..., a_n) \Leftrightarrow
i_1(a_1) \ \land ... \ \land i_n(a_n)$, 

$ 
\forall e \in  \eventset, \forall a_1, ..., a_n \in \tau_1~\times~...~\times~\tau_n, 
\forall a_1', ..., a_n' \in \tau_1'~\times~...~\times~\tau_n', 
\\\phantom{xxxx}c_1~\times~...~\times~c_n(e, a_1, ..., a_n, a'_1, ..., a'_n) \Leftrightarrow
c_1(e, a_1, a'_1) \ \land ... \ \land c_n(e, a_n, a'_n)$,

 $ 
\forall a_1, ..., a_n\in \tau_1~\times~...~\times~\tau_n, 
\\\phantom{xxxx}p_1~\times~...~\times~p_n(a_1, ..., a_n, a'_1, ..., a'_n) \Leftrightarrow
p_1(a_1, a'_1) \ \land ... \ \land p_n(a_n, a'_n)$. 
%


\subsection{An  implementation  of \foldt\ for Mercury}

We prototyped an implementation of \foldt\ for the Mercury
programming language. After a brief presentation of Mercury and its
trace system, we describe our \foldt\ implementation.

\subsubsection{Mercury and its trace}
\label{mercury-pres}

Mercury~\cite{somogyi96} is a logic and functional programming
language.  The principal differences with Prolog are as follows.
Mercury supports functions and higher-order terms. Mercury programs are
free from side-effects; even input and output are managed in a
declarative way. Mercury strong type, mode and determinism system
allows a lot of errors to be caught at compile time, and a lot of
optimizations to be done.

The trace generated by the Mercury tracer~\cite{somogyi99} is
adapted from Byrd box model~\cite{byrd80}.  Its attributes are the
event number, the call number, the execution depth, the event type
(or port), the determinism, the procedure (defined by a module name, a
name, an arity and a mode number), the live arguments, the live
non-argument variables, and the goal path. A detailed description of
these attributes together with an example of event is given in
appendix~\ref{trace-appendix}.


\subsubsection{The \foldt\ implementation} 

\label{collect-arch}

An obvious and simple way to implement \foldt\ would be to store the
whole trace into a single list, and then to apply a \fold\ to it.
This naive implementation is highly inefficient, both in time and in
space.  It requires creating and processing a list of possibly millions
of events.
Most of the time, creating such a list is simply not feasible because
of memory limitations.  With the current Mercury trace system,
several millions of events are generated each second, each event
requiring several bytes.
To implement realistic monitors, run-time information needs to be
collected and analyzed simultaneously (on the fly), {\bf without
explicitly creating the trace.}

In order to achieve analysis on the fly, we have implemented \foldt\
by modifying the Mercury trace system, which works as follows: when a
program is compiled with tracing switched on, the generated C
code\footnote{Currently, the only Mercury back-end that has a tracer
is one that relies on a C compiler to produce its executable code.}
is instrumented with calls to the tracer (via the C function
\code{trace}).  Before the first event (resp. after the last one), a
call to an initialization C function \code{trace\_init} (resp. to a
finalization C function \code{trace\_final}) is inserted.

When the trace system is entered through either one of the functions
\code{trace}, \code{trace\_init}, or \code{trace\_final}, the very
first thing it does is to look at an environment variable that tells
whether the Mercury program has been invoked from a shell, from the
standard Mercury debugger (mdb), or from another debugger (e.g.,
Morphine~\cite{morphine-ref}).
We have added a new possible value for that environment variable
which indicates whether the program has been invoked by \foldt.
In that case, the \code{trace\_init} function dynamically links the
Mercury program under execution with the object file that contains the
object code of \filter, \initialize, and \postpro.
Dynamically linking the program to its monitor is very convenient
because neither the program nor the monitor need to be recompiled.

Once the monitor object file has been linked with the
program, the C function \code{trace\_init} can call the
procedure \initialize\ to set the value of a global variable
\code{accumulator\_variable} (of type $\tau$).
At each event, the C function \code{trace} calls the 
procedure \filter\ which updates \code{accumulator\_variable}.
If \filter\ fails or if the last event is reached, the C
function \code{trace\_final} calls the procedure \postpro\ 
with \code{accumulator\_variable} and returns the new value of this
accumulator (now of type $\tau'$).


\subsection{The current user interface of \foldt\ for Mercury}

In this Section, we first describe what the user needs to do in order
to define a monitor with \foldt. 
Then, we show how this monitor can be invoked.

\subsubsection{Defining monitors}
\label{user-interface-section}

\begin{figure}
\hrule
\vspace{0.3cm}
\begin{listing}[1]{1}
    :- type accumulator_type == < A Mercury type >.

    initialize(Accumulator) :-
            < Mercury goals which initialize the accumulator >.

    collect(Event, AccumulatorIn, AccumulatorOut) :-
           < Mercury goals which update the accumulator >.

    :- type collected_type == < A Mercury type >.

    post_process(Accumulator, FoldtResult) :-
           < Mercury goals which post-process the accumulator >.
\end{listing}
\hrule

\caption{What the user needs to define to use \foldt\  \label{use-collect}}
\end{figure}


We chose Mercury to be the language in which users define the \foldt\ 
monitors to monitor Mercury programs.  As a matter of fact, it could
have been any other language that has an interface with C, since the
trace system of Mercury is written in~C. The choice of Mercury,
however, is quite natural; people who want to monitor Mercury
programs are likely to be Mercury programmers.

The items users need to implement in order to define a \foldt\ 
monitor are given in Figure~\ref{use-collect}.  Lines preceded by
\code{`\%'} are comments.
First of all, since Mercury is a typed language, one first needs to
define the type of the accumulator variable
\myemph{accumulator\_type} (line~2).
Then, one needs to define \initialize\ which gives the initial value
of the accumulator, and \filter\ which updates the accumulator at
each event(line~9). Optionally, one can also define the \postpro\ predicate
which processes the last value of the accumulator. \postpro\ 
takes as input a variable of type \myemph{accumulator\_type} ($\tau$) and
outputs a variable of type \myemph{collected\_type} ($\tau'$). If
\myemph{collected\_type} is not the same type as
\myemph{accumulator\_type}, then one needs to provide its definition
too (line~13).
Types and modes of predicates \initialize, \filter\, and \postpro\ 
should be consistent with the following  Mercury declarations:

\begin{alltt}
 :- pred \initialize(accumulator\_type::out) is det.  
 :- pred \filter(event::in, accumulator\_type::in,
         accumulator\_type::out) is semidet.
 :- pred \postpro(accumulator\_type::in, collected\_type::out) 
         is det.
\end{alltt} 

\noindent
These declarations state that \initialize\ is a deterministic
predicate (\code{is det}), namely it succeeds exactly once, and it
outputs a variable of type \code{accumulator\_type}; \filter\ is 
a semi-deterministic predicate, namely it succeeds at most once, and
it takes as input an event and an accumulator.  If \filter\ fails,
the monitoring process stops at the current event.  This can be very
useful, for example to stop the monitoring process before the end of
the execution if the collecting data is too large, or to collect data
part by part (e.g., collecting the information by slices of 10000
events). This also allows \foldt\ to operate over non-terminating
executions.

The type \code{event} is a structure made of all the event
attributes.  To access these attributes, we provide specific
functions which types and modes are:
\\
``\verb+:- func <attribute_name>(event::in) = <attribute_type>::out.+'',
\\
which takes an event and returns the event attribute corresponding to
its name.
%
For example, the function call \code{depth(Event)} returns the depth
of \code{Event}.  The complete list of attribute names is given
in Appendix~\ref{trace-appendix}.

Figure~\ref{count-call-collect} shows an example of monitor that
counts the number of predicate invocations (calls) that occur during a
program execution.
We first import library module \code{int} (line~1) to be able to
manipulate integers.  Predicate \initialize\ initializes the
accumulator to `0' (line~3). Then, for every execution event,
\filter\ increments the counter if the event port is \code{call}, and
leaves it unchanged otherwise (line~5). Since \filter\ can never
fail here, the calls to \filter\ proceed until the last event of the
execution is reached.

Note that those five lines of code constitute \emph{all the necessary
lines} for this monitor to be run.  For the sake of conciseness, in
the following figures containing monitors, we sometimes omit the
module importation directives as well as the type of the accumulator
when the context makes them clear.

\begin{figure}
\hrule
\vspace{0.3cm}
\begin{listing}[1]{1}
:- import_module int.
:- type accumulator_type == int.
initialize(0).
collect(Event, C0, C) :-
  if port(Event) = call then C = C0+1 else C = C0.
\end{listing}
\hrule

\caption{\code{count\_call}, a monitor that counts the  number of calls using \collect 
\label{count-call-collect}}
\end{figure}


\subsubsection{Invoking \foldt}
\label{calling-monitors}
\label{invoke-fold}

Currently, \foldt\ can be invoked from a Prolog query loop
interpreter.  We could not use Mercury for that purpose because there
is no Mercury interpreter yet.

We have implemented a Prolog predicate named \code{run\_mercury},
which takes a Mercury program call as argument, and which forks a process
in which this Mercury program runs in coroutining with the Prolog
process. The two processes communicate via sockets. When the
first event of the Mercury program is reached, the hand is given to
the Prolog process which waits for a \foldt\ query.

The command \foldt\ has two arguments; the first one should contain
the name of the file defining the monitor to be run; the second one
is a variable that will be unified with the result of the monitor.
When \foldt\ is invoked, (1) the file containing the monitor is used
to automatically produce a Mercury module named \code{foldt.m} (by
adding the declarations of \initialize, \filter, and \postpro, as
well as the definitions of the \code{event} type and the attribute
accessing functions); (2) \code{foldt.m} is compiled, producing the
object file \code{foldt.o}; (3) \code{foldt.o} is dynamically linked
with the Mercury program under coroutining.
Of course, steps (1) and (2) are only performed if the file
containing the monitor is newer than the object file \code{foldt.o}.

A monitor stops either because the end of the execution is
reached, or because the \code{collect} predicate failed; in the
latter case, the current event (i.e., the event the next query will
start at) is the one occurring immediately after the event where
\code{collect} failed.

\begin{figure}
\hrule
\vspace{0.3cm}
\begin{alltt}
  [morphine]: run_mercury(queens), foldt(count_call, Result).
                                      {\it A 5 queens solution is  [1, 3, 5, 2, 4]}
       Last event of queens is reached
       Result = 146     More? (;) 
  [morphine]: 

\end{alltt}
\hrule
\caption{Invoking \foldt\ monitor of Figure~\ref{count-call-collect} from an interpreter}
\label{morphine-session}
\end{figure}

A possible session for invoking the monitor of
Figure~\ref{count-call-collect} is given in
Figure~\ref{morphine-session}.  
At the right-hand side of the \codec{[morphine]:} prompt, there are the
characters typed in by a user.  The line in italic is output by the
Mercury program; all the other lines are output by the Prolog process.
We can therefore see that the program \code{queens} (which solves the
$5$ queens problem, cf Appendix~\ref{queens-prog-appendix}) produces
146 procedure calls.

\subsubsection{Illustration of the advantage of calling \foldt\ from a Prolog query loop}
\label{part-by-part-analysis-section}

Being able to call \foldt\ from a Prolog interpreter loop enables
users to write scripts that control several \foldt\ invocations.
Figures~\ref{depth-monitor} and~\ref{depth-session} illustrate this.
The monitor of Figure~\ref{depth-monitor} computes the maximal depth for
the next 500 events.  In the session of Figure~\ref{depth-session}, a
user (via the \code{[user].} directive) defines the predicate
\code{print\_max\_depth} that calls the monitor of
Figure~\ref{depth-monitor} and prints its result in loop until the
end of the execution is reached. This is useful for example for a
program that runs out of stack space to check whether this is due to
a very deep execution and to know at which events this occurs.

Note that the fact that the monitor is dynamically linked with the
monitored program has an interesting side-effect here: one can change
the monitor during the \foldt\ query resolution (by modifying the file
where this monitor is defined).  Indeed, in our example, one
could change the interval within which the maximal depth is searched
from 500 to 100. The monitor would be (automatically) recompiled, but
the \foldt\ query would not need to be killed and rerun. This can be
very helpful to monitor a program that runs permanently; the monitored
program is simply suspended while the monitor is recompiled.

\begin{figure}
\hrule
\vspace{0.3cm}
\begin{listing}[1]{1}
initialize(acc(0, 0)). 
collect(Event, acc(N0, D0), acc(N0+1, max(D0, depth(Event)))) :-
        N0 < 500.  

\end{listing}
\hrule
\caption{Monitor that computes the maximal execution depth by interval of 500 events}
\label{depth-monitor}
\end{figure}

\begin{figure}
\hrule
\vspace{0.3cm}
\begin{alltt}
  [morphine]: [user].
    print_max_depth :-
           foldt(max_depth, acc(_, MaxDepth)), 
           print("The maximal depth is "), print(MaxDepth), nl, 
           print_max_depth. 
  ^D
  [morphine]: run_mercury(qsort), print_max_depth.

    {\it The maximal depth is 54}
    {\it The maximal depth is 28}
    {\it The maximal depth is 50}
                                           {\it [0, 2, 4, 6, 7, 8, ..., 94, 95, 99, 99]}
     Last event of qsort is reached
    {\it The maximal depth is 53}
  [morphine]: 
\end{alltt}
\hrule
\caption{A possible session using the monitor of Figure~\ref{depth-monitor} }
\label{depth-session}
\end{figure}

As a matter of fact (as the prompt suggests), the Prolog query loop
that we use is  Morphine~\cite{morphine-ref}, an extensible
debugger for Mercury ``\`a la Opium''~\cite{ducasse99}.
The basic idea of Morphine is to build on top of a Prolog query loop
a few coroutining primitives connected to the trace system
(like \foldt). Those primitives let one implement all
classical debugger commands as efficiently as their hand-written
counter-parts; the advantage is, of course, that they let users
implement more commands than the usual hard-coded ones, fitting
their own needs.

Invoking \foldt\ from a debugger has a further advantage; it makes it
very easy to call a monitor during a debugging session, and vice
versa. Indeed, some monitors are very useful for  understanding program runtime
behavior, and therefore can be seen as debugging tools.

\section{Applications}
\label{collect-app}

In this section, we describe various execution monitors that can be
implemented with \foldt.  We first give monitors which compute three
different execution profiles: number of events at each port, number of
goal invocations at each depth, and sets of solutions.  Then, we
describe monitors that produce two types of execution graphs: dynamic
control flow graph and dynamic call graph.  Finally, we introduce two
test coverage criteria for logic programs, and we give the monitors
that measure them.

\subsection{Execution profiles}

\label{profiles}
\label{collect-monitoring}

\subsubsection{Counting the number of events at each port}
\label{event_count_section}

\begin{figure}
\begin{small}
\hrule
\vspace{0.3cm}
\begin{listing}[1]{1}
:- import_module int, array.   
:- type accumulator_type == array(int).

:- mode acc_in  :: array_di.
:- mode acc_out :: array_uo. 

initialize(Array) :-
    init(5, 0, Array).

collect(Event, Array0, Array) :-
    Port = port(Event),
    port_to_int(Port, IntPort),
    lookup(Array0, IntPort, N),
    set(Array0, IntPort, N+1, Array).

:- pred port_to_int(port::in, int::out) is det.
port_to_int(Port, Number) :-
    ( if    Port = call then Number = 0
    else if Port = exit then Number = 1
    else if Port = redo then Number = 2
    else if Port = fail then Number = 3
    else  Number = 4 ).
\end{listing}
%
%
%
%
\hrule

\caption{A monitor that counts the number of events
at each port} 
\label{statistic-collect}
\end{small}
\end{figure}

In Figure~\ref{count-call-collect}, we have given a monitor that
counts the number of goal invocations. Figure~\ref{statistic-collect}
shows how to extend this monitor to count the number of events at each
port.  We need 5 counters that we store in an array.
In the current implementation of \foldt, the default mode of the
second and third argument of \filter, respectively equal to \code{in}
and \code{out}, can be overridden; here, we override them with 
 \code{array\_di} and \code{array\_uo} (lines~4 and~5). Modes
\code{array\_di} and \code{array\_uo} are special modes that allow
arrays to be destructively updated.  Predicate \initialize\
creates an array \code{Array} of size 5 with each element initialized
to 0 (line~8).  Predicate \filter\ extracts the port from the
current event (line~11) and converts it to an integer
(line~12)\footnote{As a matter of fact, there are more ports than the
ones handled by {\bf port\_to\_int/2} in
Figure~\ref{statistic-collect} (cf
Appendix~\ref{trace-appendix}); we ignore them here for the sake of
conciseness.}. This integer is used as an index to get
(\code{lookup/3}) and set (\code{set/4}) array values. The goal
\code{lookup(Array0, IntPort, N)} returns in \code{N} the
\code{IntPort}$^{th}$ element of
\code{Array0}. The goal \code{set(Array0, IntPort, N+1, Array)} sets
the value \code{N+1} in the \code{IntPort}$^{th}$ element of
\code{Array0} and returns the resulting array in \code{Array}.

\subsubsection{Counting the number of calls at each depth}
\label{histogramme-collect} 
\label{histogramme-collect-section} 

\begin{figure}
\begin{small}
\hrule
\vspace{0.3cm}
\begin{listing}[1]{1}

initialize(Acc) :-
        init(32, 0, Acc).

collect(Event, Acc0, Acc) :-
    ( if port(Event) = call then
        Depth = depth(Event),
        ( if semidet_lookup(Acc0, Depth, N) then
            set(Acc0, Depth, N+1, Acc)
        else
            size(Acc0, Size),
        resize(Acc0, Size*2, 0, Acc1),
        set(Acc1, Depth, 1, Acc) 
        )
    else
        Acc = Acc0 
    ).
\end{listing}
\hrule

\caption{A monitor that counts the number of calls
at each depth} 
\label{histogram-collect}
\end{small}
\end{figure}

Figure \ref{histogram-collect} implements a monitor that counts the
number of calls at each depth. Predicate \initialize\ creates
an array of size 32 with each element initialized to 0 (line~4).  At
call events (line~7), predicate \filter\ extracts the depth
from the current event (line~8) and increments the corresponding
counter (lines~10 and~14). Whenever the upper bound of the array is
reached, i.e., whenever \code{semidet\_lookup/4} fails (line~9), the
size of the array is doubled (lines~13).

\subsubsection{Collecting solutions}
\label{collecting-solution-section}

\begin{figure}
\begin{small}
\hrule
\vspace{0.3cm}
\begin{listing}[1]{1}
:- type solution ---> proc_name/arguments.
:- type accumulator_type == list(solution).

initialize([]).

collect(Event, AccIn, AccOut) :-
    ( if
        port(Event) = exit,
        Solution = proc_name(Event)/arguments(Event),
        not(member(Solution, AccIn))
      then
        AccOut = [Solution | AccIn]
      else
        AccOut = AccIn
    ).
\end{listing}
\hrule

\caption{A monitor that collects all the solutions} 
\label{solutions-collect}
\end{small}
\end{figure}

The monitor of Figure~\ref{solutions-collect} collects the solutions
produced during the execution.  We define the type \code{solution} as
a pair containing a procedure and a list of arguments (line~1). The
collected variable is a list of \code{solution}s (line~2), which is
initialized to the empty list (line~4). If the current port is
\code{exit} (line~8) and if the current solution has not been already
produced (lines~9,10), then the current solution is added to the list
of already collected solutions (line~12).  

Note that for large programs, it would be better to use a table
from predicate names to set of solutions instead of lists.


\subsection{Graphical abstract views}
\label{graphes}

Other execution abstract views that are widely used and very
useful in terms of program understanding are given in
terms of graphs.  In the following, we show how to implement monitors
that generate graphical abstractions of program executions such as
control flow graphs and dynamic call graphs.
We illustrate the use of these monitors by applying them to the $5$
queens program given in Appendix~\ref{queens-prog-appendix}.  This
$100$ line program generates $698$ events for a board of $5 \times
5$.  In this article, we use the graph drawing tool
\code{dot}~\cite{koutsofios91}. More elaborated visualization tools
such as in~\cite{stasko98} would be desirable, especially for large
executions. This is, however, beyond the scope of this article.

\subsubsection{Dynamic control flow graphs}
\label{cfg-section}

\begin{figure} 
\begin{center} 
\hrule 
\vspace{0.3cm}
\includegraphics[width=4.5cm]{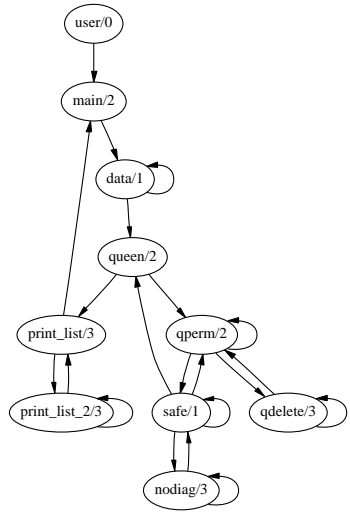}
\vspace{0.3cm}
\hrule
\caption{The dynamic control flow graph of 5 queens }
\label{cfg-eps}
 
\end{center}
\end{figure}

We define the \emph{dynamic control flow graph} of a logic program
execution as the directed graph where nodes are predicates of the
program, and arcs indicate that the program control flow went from
the origin to the destination node.  The dynamic control flow graph
of the 5~queens program is given in Figure~\ref{cfg-eps}. We can see,
for example, that, during the program execution, the control moves
from predicate \code{main/2} to predicate \code{data/1}, from
predicate \code{data/1} to predicate \code{data/1} and predicate
\code{queen/2}. Note that such a graph (or variants of it) is
primarily useful for tools and only secondarily for humans.

\begin{figure}
\begin{small}
\hrule
\vspace{0.3cm}
\begin{listing}[1]{1}
:- type predicate ---> proc_name/arity.
:- type arc ---> arc(predicate, predicate).
:- type graph == set(arc).
:- type accumulator_type ---> collected_type(predicate, graph).

initialize(collected_type("user"/0, set__init)).

collect(Event, Acc0, Acc) :-
    Port = port(Event),
    ( if (Port = call ; Port = exit ; Port = fail ; Port = redo) then
         Acc0 = collected_type(PreviousPred, Graph0),
         CurrentPred = proc_name(Event) / proc_arity(Event),
         Arc = arc(PreviousPred, CurrentPred),
         set__insert(Graph0, Arc, Graph),
         Acc = collected_type(CurrentPred, Graph)
      else    
         Acc = Acc0 
    ).
\end{listing}
\hrule
\caption{A monitor that calculates dynamic control flow graphs}
\label{cfg-source}
\end{small}
\end{figure}

\noindent

An implementation of a monitor that produces such a graph is given in
Figure~\ref{cfg-source}.
Graphs are encoded by a set of arcs, and arcs are terms composed of
two predicates (lines 1 to 3).  The collecting variable is composed of
a predicate and a graph (line~4); the predicate is used to remember
the previous node.  The collecting variable is initialized with the
fake predicate \code{user/0}, and the empty graph (line~6). At
\code{call}, \code{exit}, \code{redo}, and \code{fail} events
(line~10), we insert in the graph an arc from the previous predicate
to the current one (lines~11 to 14).

\begin{figure} 
\begin{center} 
\hrule 
\vspace{0.3cm}
\includegraphics[width=7cm]{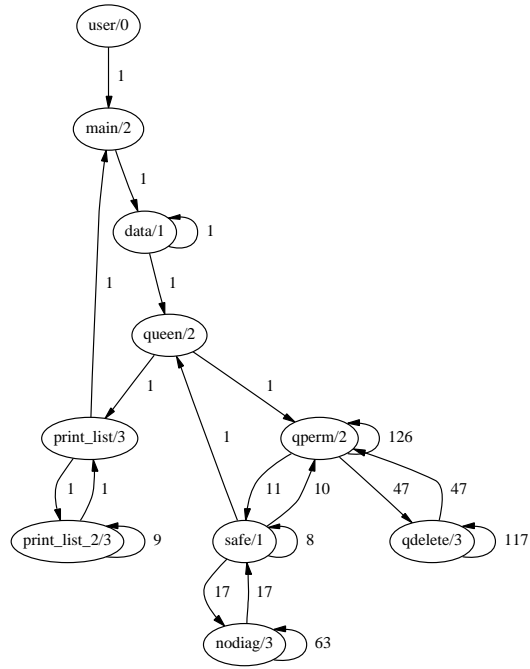}
\vspace{0.3cm}
\hrule
\caption{The dynamic {\bf control flow graph} of 5 queens annotated with counters}
\label{cfg-cpt}
\end{center}
\end{figure}

Note that in our definition of dynamic control flow graph, the number
of times each arc is traversed is not given. Even if the control
goes between two nodes several times, only one arc is represented.
One can imagine a variant where, for example, arcs are labeled by a
counter; one just needs to use multi-sets instead of sets.
The result of such a variant applied to the 5 queens program is
displayed Figure~\ref{cfg-cpt}.
Note that here, the queens program was linked with a version of the
library that has been compiled without trace information. This is the
reason why one should not be surprised not to see any call to,
e.g, \code{io\_\_write\_string/3} in this figure.


\subsubsection{Dynamic call graphs}
\label{dcg-section}

\begin{figure}
\begin{center} 
\hrule
\vspace{0.3cm}
\includegraphics[width=6cm]{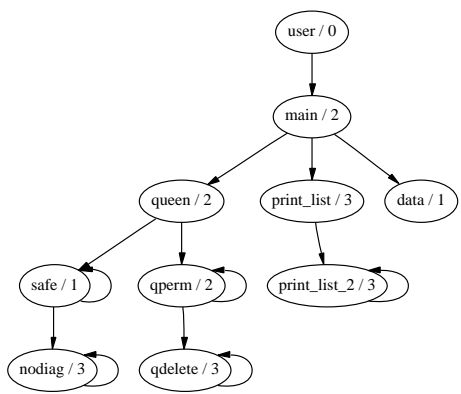}
\vspace{0.3cm}
\hrule
\caption{The dynamic {\bf call graph} of 5 queens}
 \label{dcg-eps}
\end{center}
\end{figure}

A \emph{static call graph} of a program is a graph where the nodes are
labeled by the predicates of the program, and where arcs between nodes indicate
potential predicate calls.  We define the \emph{dynamic call graph} of
a logic program execution as the sub-graph of the (static) call graph
composed of the arcs and nodes that have actually been traversed
during the execution.  For example, in Figure~\ref{dcg-eps}, we can
see that predicate \code{main/2} calls predicates \code{data/1},
\code{queen/2}, and \code{print\_list/2}.  In this particular example,
the static and dynamic call graphs are identical.

\begin{figure}
\begin{small}
\hrule
\vspace{0.3cm}
\begin{listing}[1]{1}
:- type accumulator_type ---> ct(stack(predicate), graph).

initialize(ct(Stack, set__init)) :-
      stack__push(stack__init, "user"/0, Stack).

collect(Event, ct(Stack0, Graph0), Acc) :-
      Port = port(Event),
      CurrentPred = proc_name(Event) / proc_arity(Event),
      update_call_stack(Port, CurrentPred, Stack0, Stack),
      ( if Port = call then
            PreviousPred = stack__top_det(Stack0),
            set__insert(Graph0, arc(PreviousPred, CurrentPred), Graph),
            Acc = ct(Stack, Graph)
        else
            Acc = ct(Stack, Graph0) ).

:- pred update_call_stack(trace_port_type::in, predicate::in, 
      stack(predicate)::in, stack(predicate)::out) is det.
update_call_stack(Port, CurrentPred, Stack0, Stack) :-
      ( if ( Port = call ; Port = redo ) then
            stack__push(Stack0, CurrentPred, Stack)
        else if ( Port = fail ; Port = exit ; Port = exception ) then
            stack__pop_det(Stack0, _, Stack)
        else 
            Stack = Stack0 ).
\end{listing}
\hrule

\caption{A monitor that computes dynamic call graphs}
\label{dcg-source}
\end{small}
\end{figure}

An implementation of a monitor that builds the  dynamic call graphs  is
given in Figure~\ref{dcg-source}. 
In order to define this monitor, we use the same data structures as
for the previous one, except that we replace the last traversed
predicate by the whole call stack in the collected variable type
(line~2).  This stack is necessary in order to be able to get the
direct ancestor of the current predicate.  The set of arcs is initialized to
the empty set (lines~4) and the stack is initialized to a stack that
contains a fake node \code{user/0} (line~5).  In order to construct
the set of arcs, we insert at call events an arc from the previous
predicate to the current one (line~12).  The call stack is maintained
on the fly by the \code{update\_call\_stack/4} predicate; the current
predicate is pushed onto the stack at \code{call} and \code{redo}
events (line~22), and popped at \code{exit}, \code{fail}, and
\code{exception} events (line~24).
The result of the execution of this monitor applied to
the \code{5 queens} program is displayed in Figure~\ref{dcg-eps}.

Note that the call stack is actually available in the Mercury trace.
We have intentionally not use it here for didactical purpose in order
to demonstrate how this information can easily (but not cheaply!) be
reconstructed on the fly.

\subsection{Test coverage}

\label{test}
\label{tests}

In this section, we define two notions of test coverage for logic
programs, and we show how to measure the corresponding coverage rate
of Mercury program executions using the \foldt\ primitive.  The aim
here is not to provide the ultimate definition of test coverage for
logic programs, but rather to propose two possible definitions, and to
show how the corresponding coverage rate measurements can be quickly
prototyped. As a consequence, the proposed monitors cannot pretend to
be optimal either in functionality, or in implementation.

\subsubsection{Test coverage and logic programs}

The aim of test coverage is to assess the quality of a test suite.  In
particular, it helps to decide whether it is necessary to generate
more test cases or not. For a given coverage criterion, one can decide
to stop testing when a certain percentage of coverage is reached.

The usual criterion used for imperative languages are
\emph{instruction} and \emph{branch} criteria \cite{beizer90}.  The
\emph{instruction coverage rate} achieved by a test suite is the
percentage of instructions that have been executed. The \emph{branch
  coverage rate} achieved by a test suite is the percentage of
branches that have been traversed during its execution.

One of the weaknesses of instruction and branch coverages is due to
Boolean expressions.  The problem occurs when a Boolean expression is
composed by more than one atomic instruction: it may be that a test
suite covers each value of the whole condition without covering all
values of each atomic part of the condition.  For example, consider
the condition `A or B' and a test suite where the two cases `$A=true$,
$B=false$' and `$A=false$, $B=false$' are covered. In that case, every
branch and every instruction is exercised, and nevertheless, B never
succeeded. If B is erroneous, even $100$\% instruction and branch
coverage will miss it. 
Whereas in imperative programs, you get conditional branches only in
the conditions of if-then-else and loops, in logic programs you get
them at every unification and call (whose determinism allows
failure); therefore this issue is crucial for logic programs.

\subsubsection{Predicate coverage}
\label{predicate-coverage-section}

In order to address the above problem, we need a coverage criterion
that checks that each single predicate defined in the tested program
succeeds and fails a given number of times.  But we do not want to
expect every predicate to fail because some, like printing
predicates, are intrinsically deterministic.
Therefore, we want a criterion that allows the test designer to
specify how many times a predicate should succeed and fail.
Therefore we define a \emph{predicate criterion} as a pair composed
of a predicate and a list of \code{exit} and \code{fail}.
In the case of Mercury, we can take advantage of the determinism
declaration to automatically determine if a predicate should succeed
and fail. Here is an example of predicate criterion that can be
automatically defined according to the determinism declaration of each
predicate:\\


\begin{center}
\begin{tabular}{ll}
`det' predicates: & 1 success\\ 
`semidet' predicates: & 1 success, 1 failure\\ 
`multi' predicates: & 2 successes\\ 
`nondet' predicates: & 2 successes, 1 failure\\ 
\end{tabular}
\end{center}

~\\
 
Then, we define the \emph{predicate coverage rate of a logic program
  test suite} as the percentage of program predicate criteria that
are covered during the execution of the suite.  To compute that rate,
one just needs to look at \code{exit} and \code{fail} events to see
which criterion is covered.

Figure~\ref{couv-pred} shows a \foldt\ monitor that measures the
predicate coverage rate of the \code{queens} program.
The accumulator variable is a table (\code{map}) from procedure
name to predicate criterion (line~2). 
A predicate criterion is represented by a list of \code{exit} and
\code{fail} events; the type \code{pred\_crit} also contains a list
of call numbers (line~1), initially empty (lines~6 to~14). They are
used to remember encountered exits (lines~37 to~43). Indeed, if an
execution produces two \code{exit} events for a predicate, it does
not mean that a given call of this predicate has succeeded twice; it
can be due to another call, for example recursive. Hence, we can
remove an \code{exit} atom from the list of ports to be covered only
if, either it is the first time the predicate exits (line~24), or if
the current call number has been encountered before (line~28).
Symmetrically, since all multi and nondet predicates that are called
end up with a \code{fail} event, a failure can be considered as
covered only if no \code{exit} events occurred for the current call
number before (lines~24 and~30).

The current distribution of Morphine\footnote{cf the file {\tt
    extras/morphine/source/generate\_pred\_cov.m} and the {\tt
    pred\_cov} Morphine command (both in {\tt extras.tgz} 
  available on the Mercury ftp and web sites).}
have support to automatically generate such monitors, run them, and
compute the coverage percentage. Monitors are generated by parsing
the source files in order to get the procedure determinisms that are
necessary to be able to produce the right number of \code{exit} and
\code{fail} atoms.

\begin{figure}
\hrule
\vspace{0.3cm}
\begin{small}
\begin{listing}[1]{1}
:- type pred_crit ---> pc(list(call_number), list(port)).
:- type accumulator_type == map(proc_name, pred_crit).

initialize(Map) :-
  map__init(Map0),
  map__det_insert(Map0, "main",        pc([], [exit]), Map1),
  map__det_insert(Map1, "data",        pc([], [exit]), Map2),
  map__det_insert(Map2, "print_list",  pc([], [exit]), Map3),
  map__det_insert(Map3, "print_list_2",pc([], [exit]), Map4),
  map__det_insert(Map4, "safe",   pc([], [exit,fail]), Map5),
  map__det_insert(Map5, "nodiag", pc([], [exit,fail]), Map6),
  map__det_insert(Map6, "qperm",  pc([], [exit,exit,fail]), Map7),
  map__det_insert(Map7, "qdelete",pc([], [exit,exit,fail]), Map8),
  map__det_insert(Map8, "queen",  pc([], [exit,exit,fail]), Map).

collect(Event, Map0, Map) :-
  Port = port(Event), Proc = proc_name(Event), CallN = call(Event),
  ( if 
      (Port = exit ; Port = fail), pc(CNL0, PL0) = map__search(Map0, Proc) 
    then
    ( if 
          CNL0 = [] 
        then 
          remove_port(Port, PL0, PL) 
        else if 
          member(CallN, CNL0) 
        then 
          (if Port = exit then remove_port(exit, PL0, PL) else PL = PL0)
        else 
          (if Port = exit then PL = PL0 else remove_port(fail, PL0, PL))
      ),
      ( if 
          PL = [] 
        then 
          map__delete(Map0, Proc, Map) 
        else 
          ( if 
              Port = exit, not member(CallN, CNL0)
            then 
              CNL = [CallN | CNL0]
            else 
              CNL = CNL0 
          ),
          map__update(Map0, Proc, pc(CNL, PL), Map))
    else
      Map = Map0 ).

:- pred remove_port(port::in, list(port)::in, list(port)::out) is det.
remove_port(Port, L0, L) :-
  if list__delete_first(L0, Port, L1) then L = L1 else L = L0.

:- type collected_type == assoc_list(proc_name, pred_crit).
post_process(Map, AssocList) :- map__to_assoc_list(Map, AssocList).
\end{listing}
\end{small}
\hrule
\caption{A monitor that measures the predicate coverage rate of the n queens 
program} 
\label{couv-pred}
\end{figure}
 
\subsubsection{Call site coverage}
\label{call-site-coverage}

The previous coverage criterion only checks that at least one exit for
each predicate is covered.  The problem is that $100$\% predicate
coverage does not imply $100$\% instruction nor branch coverage.
To ensure $100$\% instruction and branch coverage, we need a criterion
that ensures that every \emph{predicate invocation} in the program
succeeds and fails.
Hence we need a definition attached to call sites (or goals) and not
only to predicates.  To achieve that, we just need, for example, to take
advantage of line numbers and having a table from procedure
names and line numbers to list of ports as accumulator.
The monitor that measures call site coverage rate of Mercury program
executions is therefore roughly the same as the one given in
Figure~\ref{couv-pred}, where the accumulator type becomes:

\begin{verbatim}
:- type proc ---> p(declared_module_name, proc_name, line_number).
:- type call_site_crit ---> csc(list(call_number), list(port)).
:- type accumulator_type == map(proc, call_site_crit).
\end{verbatim}

Here again, such monitors are generated automatically parsing the
 source files\footnote{cf the file {\tt
    extras/morphine/source/generate\_call\_site\_cov.m} and the {\tt
    call\_site\_cov} Morphine command.}.



\section{Experimental Evaluation}
\label{performance-section}

In the previous section we have shown the flexibility and power of the
\foldt\ primitive. The aim of this section is to assess the
performance of the current \foldt\ implementation.
When executing a monitor, some time is spent in the normal program
execution ($T_{prog}$); and some extra time is spent in the trace system
of Mercury ($\Delta_{trace}$), in the interface between the tracer
and the \foldt\ mechanism \footnote{The Mercury predicate
\code{collect} is called from the Mercury tracer which is written in C.} 
($\Delta_{inte}$), in the basic \foldt\ mechanism
($\Delta_{foldt}$), and also in the monitor itself
($\Delta_{<monitor>}$). Hence, if we call $T$ the execution time of a
monitored program, we have:
\begin{center}
$T = T_{prog} + \Delta_{trace} + \Delta_{inte} + \Delta_{foldt} +
\Delta_{<monitor>}$
\end{center}

In the following, we measure:
$ T_{prog}$ ,
$ T_{trace} = T_{prog} + \Delta_{trace} $ , 
$ T_{inte} = T_{prog} + \Delta_{trace} + \Delta_{inte}$ , 
and $ T_{foldt} = T_{prog} + \Delta_{trace} + \Delta_{inte} + \Delta_{foldt}$.
%

We compare $T_{trace}$, $ T_{inte}$, and $T_{foldt}$ against
$T_{prog}$. We will therefore compute the following ratios:
\begin{center}
$R_{t} = T_{trace}/T_{prog}$, ~~
$R_{i}= T_{inte}/T_{prog}$ ~~and~~
$R_{f} = T_{foldt}/T_{prog}$
\end{center}

\subsection{Methodology} 

\paragraph{\bf Hardware and software.}
The measurements given in the following show the results of
experiments run on a DELL inspiron 7500, with a 433~MHz Celeron,
192~Mb of RAM, running under the Linux~2.2.14 operating system. The
machine was very lightly loaded; no X server, no network, simply the
basic operating system and a Prolog process in a console to run the
measurement scripts. The Prolog system is Eclipse
4.1~\cite{eclipse99}. The Mercury compiler is a stable snapshot of 14
June 2001\footnote{The last official release is numbered 0.10.1}. The
results are consistent with experiments run on a SUN Sparc Enterprise
250 (2 x UltraSPARC-II, 296MHz, 512 Mb of RAM) running Solaris 2.7
(figures not given here).

\paragraph{\bf Time measuring command.}
In order to measure the program execution times, we use the
\code{benchmark\_det} predicate of the \code{benchmarking.m} Mercury
standard library. This predicate repeats the body of a program any
given number of times. This is very important for small programs, as
the startup cost very often dominates the execution cost.  In the
following experiments, each program is re-executed until it runs at
least for 20 seconds.
Each experiment has been done five times, and the deviation was
smaller than 1 \%.

\paragraph{\bf Monitored programs.}
The monitored programs are the Mercury benchmark suite\footnote{The
source code of this benchmark suite can be found on the Mercury ftp
site
\code{ftp://www.mercury.cs.mu.oz.au/pub/mercury/mercury-tests-*.tar.gz}},
composed of programs adapted from the Prolog benchmark suite
of~\cite{vanroy92}.  In order to have a wider range of execution
sizes, we also measure \code{n~queens} for n=10,11, as well as
\code{mastermind}, a 1100 lines Mercury program which solves a
mastermind game\footnote{The full source code of the Mercury
mastermind program can be found at first author web site}.

\paragraph{\bf Mercury compilation grades.}
In the following, the compilation grade $\code{g_{nt}}$ refers to
Mercury modules compiled with the command {\tt mmc {-}{-}grade
asm\_fast.gc.picreg}, which means that no trace event is generated. It
is the grade used to measure the plain execution time of programs
($T_{prog}$).

The compilation grade $\code{g_t}$ refers to Mercury modules compiled
with  {\tt mmc {-}{-}grade asm\_fast.gc.picreg {-}{-}trace
deep --trace-optimized}, which \\means that all events related to
all the predicates of the module, except library predicates, are
generated.  This grade is used in the following to measure the time
spent in the basic trace system ($T_{trace}$), the time spent in the
interface between the basic tracer and the \foldt\ mechanism
($T_{inte}$), the time spent in the basic \foldt\ mechanism
($T_{foldt}$) and the time spent in the monitors ($T_{<monitor>}$).

\paragraph{\bf Measuring $T_{prog}$.}
If the programs are compiled in grade $\code{g_{nt}}$, then their execution
does not produce any trace and their measured execution duration
($T_{measured}$) is exactly $T_{prog}$.

\begin{center}
$T_{prog} = T_{measured}$ ~~if compilation grade is $\code{g_{nt}}$
\end{center}

\paragraph{\bf Measuring $T_{trace}$.}
If the programs are compiled in grade $\code{g_{t}}$, then their execution
calls the Mercury tracer. In order to measure the cost of the basic
tracer, we have to ensure that the tracer is systematically called
at each event, but that it does not do anything else than entering and
exiting the top-level switch of the trace system.

\begin{center}
$T_{trace} = T_{measured}$ ~~if compilation grade is $\code{g_{t}}$
and $\Delta_{inte} = 0, \;\; \Delta_{foldt} = 0, \;\; and \;\;\Delta_{<monitor>} = 0$
\end{center}
In order to ensure that $\Delta_{inte} = 0, \;\; \Delta_{foldt} = 0\;\; and
\;\;\Delta_{<monitor>} = 0$, we use the $\code{continue}$ command of the
Mercury tracer without specifying any break-point. Indeed, with that
command, at each event, the trace system is entered; if the event does
not correspond to one of the specified breakpoints, the normal
execution is resumed. In our measurements, as no break-point is
specified, the whole execution is traversed, and nothing is executed
but the basic tracing mechanism.

\paragraph{\bf Measuring $T_{inte}$.}

In order to measure the cost of the interface between the
basic tracer and the \foldt\ mechanism, we have to ensure that the
tracer is systematically called at each event, that it enters and
exits the top-level switch of the trace system, that it prepares the
context to call the \code{collect} predicate defined for the monitor,
but that it does not compute anything else, in particular it should
not retrieve any event attribute.

\begin{center}
$T_{inte} = T_{measured}$ ~~if compilation grade is $\code{g_{t}}$
and $\Delta_{foldt} = 0, \;\; and \;\;\Delta_{<monitor>} = 0$
\end{center}
In order to ensure that $\Delta_{foldt} = 0$, we have implemented a
degenerate \foldt\ such that no event attribute is computed (we have
replaced these computations by void values).
In order to ensure that $\Delta_{<monitor>} = 0$, we use a monitor that
does not compute anything (\code{collect(\_E, A, A).}).

\paragraph{\bf Measuring $T_{foldt}$.}
In order to measure the cost of the basic \foldt\ mechanism, we have
to ensure that  \code{collect} is called at each event for a monitor
that computes nothing.

\begin{center}
$T_{foldt} = T_{measured}$ ~~if compilation
grade is $\code{g_{t}}$ and $\Delta_{<monitor>} = 0$
\end{center}
In order to make sure that $\Delta_{<monitor>} = 0$, we 
call \foldt\ with a trivial monitor, that does not compute
anything (\code{collect(\_E, A, A).}).

Two of the current attributes are very costly to retrieve: the live
arguments and the line number.
The live arguments can be very large data structures. 
The line number corresponds to the line where the goal is called and
not where the predicate is defined. It is dynamically retrieved.
Many interesting monitors can be run without these attributes.
Indeed, for the monitors we propose in this article, only one monitor
uses the live arguments and one monitor uses the line number.
Monitors that do not use these costly attributes can disable them. 
As a consequence, the measurements of $T_{foldt}$ are made with these
two attributes disabled.



\subsection{Resulting table}
\label{measuring-result-section}

\begin{table}
\begin{center}
\caption{Cost of the  \foldt\ mechanisms on 
benchmarks. $T_{trace} = T_{prog} + \Delta_{trace}$, 
$T_{inte} = T_{prog} + \Delta_{trace} + \Delta_{inte}$,
$T_{foldt} = T_{prog} + \Delta_{trace} + \Delta_{inte} + \Delta_{foldt}$,
$R_{t}~=~T_{trace}/T_{prog}$, 
$R_{i}~=~T_{inte}/T_{prog}$,
$R_{f}~=~T_{foldt}/T_{prog}$,
$R^*_{f}~=~(T_{foldt}-\Delta_{inte})/T_{prog}$}
\renewcommand{\arraystretch}{1.4}
\setlength\tabcolsep{4pt}
\begin{footnotesize}
\begin{tabular}{l r c c c c c c c c}
\noalign{\smallskip}
\hline
\noalign{\smallskip}
program     &  events & $T_{prog}$ 
&  $T_{trace}$ &$R_{t}$  
& $T_{inte}$ & $R_{i}$ 
& $T_{foldt}$ & $R_{f}$ & $R^*_{f}$ \\
&& in $ms$ & in $ms$ && in $ms$ && in $ms$ &&\\
\hline queens-5 & 698           &  0.03 & 0.24  & 7.5 & 2     & 61.5  & 2.07   & 63   & 9.5 \\ 
\hline query    & 935           &  0.09 & 0.28  & 3   & 1.53  & 16.5  & 1.58   & 17   & 3.5 \\ 
\hline deriv    & 1,540         &  0.05 & 0.11  & 2.5 & 0.64  & 14    & 0.66   & 14.5 & 3 \\ 
\hline qsort    & 1,564         &  0.1  & 0.48  & 5   & 4.03  & 42    & 4.16   & 43.5 & 6.5 \\ 
\hline nrev     & 1,619         &  0.14 & 0.54  & 4   & 4.44  & 32.5  & 4.58   & 33.5 & 5 \\ 
\hline primes   & 2,192         &  0.21 & 0.8   & 4   & 6.23  & 30    & 6.51   & 31   & 5.5 \\ 
\hline cqueens  & 3,789         &  0.14 & 1.26  & 9.5 & 11.11 & 81    & 11.39  & 82   & 11.5 \\ 
\hline crypt    & 4,602         &  0.72 & 1.8   & 3   & 11.16 & 16    & 11.54  & 16.5 & 3.5 \\ 
\hline poly     & 79,070        &  6.44 & 29    & 4.5 & 226.2 & 35.5  & 233.4  & 36.5 & 6 \\ 
\hline tak      & 190,831       &  3.88 & 57.1  & 15  & 553.7 & 142   & 572.4  & 148  & 20 \\ 
\hline queens-10  & 4,289,986   &  257  & 1530  & 6   & 12760 & 50.5  & 13200  & 51.5 & 8 \\ 
\hline mastermind & 9,106,510   &  3630 & 6500  & 2   & 30490 & 8.5   & 31520  & 9    & 2.5 \\ 
\hline queens-11  & 32,384,320  &  2103 & 12030 & 6   & 97010 & 46.5  & 100190 & 48   & 7.5 \\ 
\hline

\end{tabular}

\end{footnotesize}
\label{perf-bench-table}
\end{center}
\end{table}

Table~\ref{perf-bench-table} illustrates the cost of the basic tracer,
the cost of the interface between the tracer and \foldt, as well as the
cost of \foldt\ on the benchmark programs described in the previous
sections.
The first column contains the names of the monitored programs. 
The second column contains the numbers of execution events generated by
the program executions compiled in grade $\code{g_t}$ (all events are
generated, except events relative to library predicates). 
The programs are sorted wrt this number of events, from \code{queens-5}, 698
events, to \code{queens-11}, more than 32 millions of events.
The third column contains the execution times of the programs compiled
without any trace information ($T_{prog}$). 
The fourth column contains the execution times of the programs compiled in
grade $\code{g_t}$ and run under the control of the tracer without tracing
anything ($T_{trace} = T_{prog} + \Delta_{trace}$). 
The fifth column contains the overhead factor of the basic trace
mechanism ($R_{t} = T_{trace} / T_{prog}$).
The sixth column contains the execution times of the programs compiled
in grade $\code{g_t}$ and run under the control of the tracer, and
where the degenerate \foldt\ is called with an empty monitor
($T_{inte} = T_{prog} + \Delta_{trace} + \Delta_{inte}$).
The seventh column contains the overhead factor of the trace and the
interface between the tracer and \foldt\ ($R_{i} = T_{inte} / T_{prog}$).
The eighth column contains the execution time of the programs compiled
in grade $\code{g_t}$ and run under the control of \foldt\ with an
empty monitor ($T_{foldt} = T_{prog} + \Delta_{trace} + \Delta_{inte}
+ \Delta_{foldt}$).
The ninth column contains the overhead factor of the trace, the
interface and the basic \foldt\ mechanism ($R_{f}= T_{foldt} /
T_{prog}$).
The tenth column contains the overhead factor of the trace and the
basic \foldt\ mechanism, with the interface cost divided
($R^*_{foldt}= (T_{foldt} - \Delta_{inte}) / T_{prog}$).
The time measurements have been rounded off two digits after the dot.
The ratios have been rounded up to the nearest half.

\subsection{Discussion}

In this section, we discuss the resulting ratios of
Table~\ref{perf-bench-table}.

\paragraph{\bf Two extremes: \code{tak} and \code{mastermind}.} The
\code{tak} program has ratios much higher than the other programs. The
tracer overhead is already 15, the interface overhead is 142
and the \foldt\ overheads are 148 and 20. This program is actually a
single predicate four times recursive. It already broke the stacks of
the reference tracer used in \cite{ducasse2000}. This code is an
extreme case to test compiler optimization capabilities. As many
optimizations, such as last call optimization, cannot be applied to
traced code, the better the compiler is, the worse debugger and
monitor ratios look. Program \code{Tak} is very untypical of real life
programs. The ratios related to \code{tak} are not taken into account
in the averages given below.

On the other hand \code{mastermind} has very low ratios. The tracer
overhead is 2, the interface overhead is 8.5 and the \foldt\
overheads are 9 and 2.5. The \code{mastermind} program uses a lot of
library predicates which are not traced\footnote{their calls and exits are
traced, but not what happens inside these calls} and monitored in
detail. This is typical of real life programs. It is encouraging
that the more realistic program has the best ratios. However, the
other benchmarks do not use untraced libraries, in order to be fair,
the ratios related to \code{mastermind} are not taken into account in
the averages given below.

In the following, averages are, thus, computed without the figures
related to \code{tak} and \code{mastermind}.

\paragraph{\bf The ratios are not correlated with the number of events.}
Program \code{queens-5}, which has only 698 events, has very bad
ratios, whereas \code{crypt}, almost five thousand events, and
\code{poly}, almost eighty thousand events, have better ratios than
the average. The same program, \code{n~queens}, run for n= 5,10 and 11,
always has comparable ratios. This seems to indicate that the
overheads depend mostly of the monitored program and is somewhat
constant for a given program.

\paragraph{\bf Overhead of the tracer.}
The  average of the tracer overhead is 5. It is a very
good ratio. Prolog tracers can easily have an overhead over 20
\cite{ducasse2000}. A low ratio for the tracer is, of course, a
good starting point to build efficient generic monitors.

\paragraph{\bf Overhead of the interface between the tracer and \foldt.}

The average of the interface overhead is 39. This is very high and it
is the main source of inefficiency of our current implementation. It
illustrates how crucial the implementation of this interface is for
efficient generic monitoring.

The problems comes from the fact that the monitor programs have to be
integrated in the tracer. 
In our particular case, the Mercury predicate \code{collect} is called
from the Mercury tracer which is written in C. In order to call
Mercury code from C with the current (low-level back-end of the)
Mercury compiler, machine registers need to be saved and restored.
Since the \code{collect} predicate is called several million times,
this has a noticeable influence on the performance.
A way to remove this overhead could be to use the new MLDS back-end of
the Mercury compiler, which generates high-level C code that does
not use machine registers; unfortunately, the trace system is not
supported for the MLDS back-end at the time of this writing.
Once the MLDS Mercury back-end is available, calling the
\code{collect} predicate will actually be compiled as a simple C
procedure call from within C code.  The overhead of the interface
between the tracer and \foldt\ should thus become smaller.

One important lesson learned from these measurements is as
follows. Whether the monitors and the tracer are implemented in the
same programming language or not, the integration of the compiled
monitors should not cost more than a procedure call. The monitors must
therefore be compiled into the same target language as the
tracer. Furthermore, no run-time verifications should be made. The
monitors should therefore have no side-effect on the traced
execution. It should thus be statically checked that monitors only
update their own (fresh) variables.

\paragraph{\bf Overhead of \foldt.}

The average of the \foldt\ overhead is 40. Most of it is due to
the overhead of the interface discussed above. Assuming that the above
fix could be done and that the interface overhead indeed becomes
negligible, we have computed an ideal ratio: $R^*_{f} =
(T_{foldt}-\Delta_{inte})/T_{prog}$.  The average of the overhead of
\foldt\ without the interface cost is 6.5. As the average of the
tracer overhead is 5, we can say that the \foldt\ overhead is
acceptable.

In the context of Mercury, this is especially true because the
developers of Mercury claim that Mercury programs executed in trace
mode are faster than the equivalent Prolog programs executed in 
optimized mode in the faster Prolog systems \cite{somogyi99}.

\paragraph{\bf Unused event attributes.} 
As already mentioned, some attributes can be very costly to
compute. When they are not needed it should be possible to disable
them.  In the current implementation of \foldt, this is already the
case for the list of live arguments and the line numbers.
The foldt overhead has been measured without the cost of the live
arguments and the line numbers. 
Some measurements, not reported here, showed that this has an impact on
the performance.

\paragraph{\bf Granularity of the instrumentation.} 
In order to measure the worst case, we have made the Mercury tracer
systematically generate all the possible events. Not all  monitors
need such a fine-grained instrumentation. For example, for the monitor
that counts the number of events at each depth, only \code{call}
events are necessary. 
When one (hard-)codes a specific monitor, one only instruments the
code where it is necessary.
As a matter of fact, the Mercury tracer enables users to specify what
type of events, if any, should be generated for a given module (the
only restriction is that, if some events are generated for a
predicate, \code{call} events must be present).  Hence, programmers
can already take advantage of this possibility to optimize their
monitor. As further work, we plan to automate this optimization,
namely, to automatically generate the appropriate compiling option
for a given monitor.
 





\paragraph{\bf Conclusion on performance.}

The cost of monitors is generally superseded by the cost of the
\foldt\ mechanism, except for time demanding monitors such as the one
that computes dynamic call graphs and coverage rates for which we have
yet another slowdown of a factor ranging from 1.5 to 5. 
Hence, our conclusion is that with a fast tracer, an interface
between the tracer and \foldt\ reduced to procedure calls, and the
possibility to disable the computation of heavy non-necessary
attributes,  generic monitoring can be efficient enough.



\section{Related work}
\label{collect-related}

\paragraph{Programmable debuggers.}
We designed 3 programmable debuggers, Opium for Prolog
\cite{ducasse99}, Coca for C \cite{ducasse99c} and Morphine for
Mercury \cite{morphine-ref}.  They are based on a Prolog query loop
plus a handful of coroutining primitives connected to the trace
system. Those primitives allow a Prolog system to communicate
with the debugged program. Opium, Coca and Morphine are full debugging
programming languages in which all classical debuggers commands can be
implemented straightforwardly and efficiently. However, 
 their set of primitives are not well suited for monitoring.  All
the monitors implemented with \foldt\ can easily be implemented with
the debugger set of primitives \cite{jahier99b}, but resulting
monitors require too many context switches and too much socket traffic
between the program and the monitor.  With programs of several million
execution events, such monitors can be four orders of magnitude slower
than their counterparts that use \foldt\ \cite{jahier2000d}.

\paragraph{Automated development of monitors.}
Jeffery et al. designed the Alamo system~\cite{templer98b,jeffery98,jeffery99},
 that aims at easing the development of monitors for C
programs.  As in our approach, their monitoring architecture is based
on event filtering, and monitors can be programmed. Their system
performs trace extraction whereas we rely on an already available
tracer; this saves us a tedious task which has already been done and
optimized. On the other hand, we do not have the full control of the
information available in the trace. Note however that, so far, we
have be able to reconstruct missing information, for example the call
stack of Figure~\ref{dcg-source}.  Moreover, to avoid code explosion,
Jeffery et al. perform part of the event filtering at compilation
time. This means that they need to recompile the program each time
they want to execute another monitor, whereas we only need to
dynamically link the monitor to the monitored program. Alamo and the
monitored program are running in coroutining, but within the same
address space.

Eustace and Srivastava developed Atom~\cite{eustace95},
a system that also aims at easing the implementation of monitors. The
difference with Alamo is that monitors are implemented with procedure
calls and global variables which is much more efficient than
coroutining.  However, the language provided by Atom is  less
expressive than the Alamo's.  Alamo and Atom have influenced the
design of \code{foldt} and we tried to take the best of both: a
full and high-level programming language implemented by procedure
calls.
The advantages of our architecture over~\cite{eustace95} and
\cite{jeffery98} are the following:

\begin{itemize}
\item A higher level interface makes the code of our monitors compact,
easy to write and read, and therefore to maintain.
Of course, this point is difficult to assess. We hope that the
numerous and various examples given in Section~\ref{collect-app}
convince the reader that it is indeed the case.
Nonetheless, we see three conjectures explaining why the code is more
compact and more elegant. Firstly, users do not have to deal with
code instrumentation directives; the instrumentation has already been
done. 
Secondly, we take advantage of the expressive power of
\fold; using high order predicates such as \fold\ for processing lists
has proven to be concise and far less prone to error than processing
lists manually. A third reason is that our monitors are written in
Mercury, which is a considerably higher level language than C or
C-like languages which are used in~\cite{eustace95,jeffery98}.

\item Provided that an event-oriented tracer exists, the \foldt\
operator is easy to implement. To implement it, the work done inside
the Mercury runtime system, which corresponds to the really technical
part, amounts to only 61 modified lines and 292 new lines of (C) code.
\end{itemize}

\paragraph{}
Kishon and al.~\cite{kishon95} use a denotational and operational
continuation semantics to formally define monitors for a simple
functional programming language. The kind of monitors they define are
profilers, debuggers, and statistic collectors. From the operational
semantics, a formal description of the monitor, and a program, they
derive an instrumented executable file that performs the specified
monitoring activity. The semantics of the original programs is
preserved. They use partial evaluation to make their monitors
reasonably efficient.
The main disadvantage with this approach is that they are rebuilding a
whole execution system from scratch, without taking advantage of
existing compilers.  We strongly believe that it is important to have
the same execution system for debugging, monitoring and  producing
the final executable program. As noted by 
\cite{brooks92}, some errors only occur in presence of optimizations,
and vice versa; some programs can only be executed in their optimized
form because of time and memory constraints; when searching for ``hot
spots'', it is better to do it as much as possible with the optimized
program as many things can be optimized away; and finally, sometimes,
the error comes from the optimizations themselves. In our setting we
can easily mix traced and non-traced code.

\paragraph{Efficient monitoring.}
Patil and Fisher \cite{patil97} address the problem of performance
monitoring by delegating the monitoring activities to a second
processor that they call a shadow processor. Their approach is very
efficient; the monitored program is practically not slowed down, but
the set of monitoring commands they propose cannot be extended.
We mentioned in the previous section that we could reduce the number
of events generated by the tracer. 
For example,
in \cite{ball92}, given a static control flow graph, algorithms can
place tracing instructions in optimal ways for computing statistics on
imperative program executions.

\section{Conclusion}
\label{collect-conclusion}

In this article we have proposed a generic monitoring framework based
on \foldt\footnote{Available in Morphine, which can be downloaded from
the Mercury ftp and web sites.}, a high-level primitive that allows
users to easily specify what they want to monitor. We illustrated it
on various examples that demonstrate its genericity and its
simplicity of use.  We defined two preliminary notions of test
coverage for logic programs and showed how to prototype coverage
rates measurements with our primitive.
Testing and monitoring tools are missing from many declarative
systems: \foldt\ allows some of these tools to be easily defined and
implemented.
Measurements showed that the performance of the primitive on the above
examples can be acceptable for executions of several million trace
events.

\paragraph{}
To sum up the advantages of our framework, we can say that it is:
\begin{itemize}
\item Easy to implement: because it is based on an existing tracer
(292 new, and 61 modified lines of codes in our current
implementation).
\item Efficient: because the trace is not stored.
\item Flexible and easy to use: as illustrated by the given
applications about execution profiles, graphical abstract displays and
test coverage.
\end{itemize}

\section*{Acknowledgments} We would like to thank Fergus Henderson
for his technical support and his many contributing ideas; Pierre
Deransart, Baudouin Le Charlier and Olivier Ridoux for fruitful
discussions; Jean-Philippe Pouzol for his comments on earlier versions
of this article.
Finally, we are grateful to the thorough anonymous reviews which have
helped us a lot improve the article.




\newpage
\appendix

\section{Mercury execution events}
\label{trace-appendix}

The Mercury trace is an adaptation of Byrd's box model~\cite{byrd80}.
In this section, we describe the Mercury execution events  that
constitute the Mercury execution trace. More information about the
Mercury tracer can be found in \cite{somogyi99}.
The different \emph{attributes} provided by the Mercury tracer are:

\begin{enumerate}
\item \emph{Chronological event number} (\code{chrono}\footnote{The
    names of the attribute accessing functions are in bold in between parentheses.}).
%
Each event has a unique event number according to its rank in the
trace. It is a counter of events.

\item  \emph{Goal invocation number} or \emph{call number} (\code{call}). 
Unlike chronological event number, several events have
the same goal invocation number.
All events related to a given goal have a unique goal number given at
invocation time.

\item  \emph{Execution depth} (\code{depth}).
%
It is the depth of the goal in the proof tree, namely the number of its
ancestor goals + 1.

\item  \emph{Event type or port} (\code{port}).
%
  We distinguish between \emph{external events} that occur at
  procedure entries and exits, which are the traditional ports
  introduced by Byrd \cite{byrd80}, and the \emph{internal events}
  which refers to what is occurring inside a procedure.
External events are:
\begin{itemize}
\item \code{call} a new goal is invoked
\item \code{exit} the current goal succeeds
\item \code{fail} the current goal fails
\item \code{redo} another solution for the current goal is asked for 
on backtracking.
\item \code{exception} the execution raises an exception
 \end{itemize}

\noindent
Internal events are:
\begin{itemize}

\item \code{disj} the execution is entering a branch of a disjunction
\item \code{switch} the execution is entering a branch of a
  switch  (a \emph{switch} is a disjunction in
    which each branch unifies a ground variable with a different
    function symbol. In that case, at most one disjunction 
    provides a solution). 
\item \code{if} the execution is entering the condition branch of an
  if-then-else
\item \code{then} the execution is entering the ``then'' branch of an
  if-then-else
\item \code{else} the execution is entering the ``else'' branch of an
  if-then-else
\item \code{first} the execution enters a C code fragment for the
  first time
\item \code{later} the execution re-enters a C code fragment
\end{itemize}

\item  \emph{Determinism} (\code{det}).
%
  It characterizes the number of potential solutions for a given
  goal. The determinism markers of Mercury are: \code{det} for
  procedures which have exactly 1 solution, \code{semidet} for those
  which have 0 or 1 solution, \code{nondet} for those which have any
  number of solutions, \code{multi} for those which have at least 1
  solution, \code{failure} for those that have no solution, and
  \code{erroneous} for those which lead to a runtime error.
  
\item \emph{Procedure} (\code{proc}).
It is defined by:
\begin{itemize}
\item \emph{a flag} telling if the procedure is a function or a
    predicate (\code{proc\_type})
\item  \emph{a definition module} (\code{def\_module}) 
\item  \emph{a declaration module} (\code{decl\_module})  
  The declaration module is the module where the user has declared
  the procedure.  The defining module is the module where the
  procedure is effectively defined from the compiler point of view.
  They may be different if the procedure has been inlined.
\item  \emph{a name} (\code{name})
\item  \emph{an arity} (\code{arity}) 
\item  \emph{a mode number} (\code{mode\_number}).

The mode number is an integer coding the mode of the procedure. When
a predicate has only one mode, the mode number of its corresponding
procedure is 0. Otherwise, the mode number is the rank in the order
of appearance of the mode declaration.  

\end{itemize}
\item \emph{List of live arguments} (\code{args}).
  A variable is \emph{live} at a given point of the execution if it
  has been instantiated and if the result of that instantiation is
  still available in the runtime system. Destructive input (\code{di}
  mode), for example, are not kept until the procedure exits.

\item  \emph{List of live Argument types} (\code{arg\_types}).

\item  \emph{List of local live variables} (\code{local\_vars}).
Some live variables are not arguments of the current procedure.

\item \emph{Goal path} (\code{goal\_path}).
  The goal path indicates in which part of the code the current
  internal event occurs. \code{if}, \code{then} and \code{else}
  branches of an \emph{if-then-else} are denoted by \code{?},
  \code{e} and \code{t} respectively; \emph{conjuncts},
  \emph{disjuncts} and \emph{switches} are denoted by \code{ci},
  \code{di} and \code{si}, where \code{i} is the conjunct (resp.
  disjunct, switch) number.  For example, if an event with goal path
  \code{[c3, e, d1]} is generated, it means that the event occurred
  in the first branch of a disjunction, which is in the else branch
  of an if-then-else, which is in the third conjunction of the
  current goal.  External events have an empty goal path.


\begin{figure}[htbp]
\begin{center}
\begin{tabular}{ll}
\hline
\verb+chrono+ & \verb+10 +\\
\verb+call+ &  \verb+6+\\
\verb+depth+ & \verb+5 +\\
\verb+port+ &  \verb+then+\\
\verb+det+ & \verb+det+\\
\verb+  proc_type+ & \verb+predicate+\\
\verb+  def_module+ & \verb+qsort+\\
\verb+  decl_module+ & \verb+qsort+\\
\verb+  name+ &  \verb+partition+\\
\verb+  arity+ &  \verb+4+\\
\verb+  mode_number+ & \verb+0+\\
\verb+arg+ &  \verb+[ [1, 2], 3, -, - ]+\\
\verb+arg_types+ &  \verb+[list(int), int, -, - ]+\\
\verb+local_vars+ &  
   \verb+[ live_var("H", 1, int), live_var("T", [2], list(int)) ]+\\
\verb+goal_path+ & \verb+[s1, c2, t]+\\
\hline
\end{tabular}
\end{center}
\caption{A Mercury trace event} 
\label{event_structure}
\end{figure}

The event structure is illustrated by Figure~\ref{event_structure}.
The displayed structure is related to an event of the execution of a
\code{qsort} program which sorts the list of integers \code{[3,~1,~2]}
using a \emph{quick sort} algorithm.
%
%
The information contained in that structure indicates that 
\code{qsort:partition/4-0}\footnote{\code{`-0'} denotes the mode number; here, \code{`0'}
 means that \code{qsort} was declared with only one mode (namely,
\code{:- mode qsort(in, in, out, out) is det}). If more than one mode
is declared, \code{`0'} denotes the first mode, \code{`1'} the
second one, etc.} is currently invoked, it is the tenth trace event being
generated, the sixth goal being invoked, and it has four ancestors
(depth is 5).  At this point, only the first two arguments of
\code{partition/4} are instantiated: the first one is bound to the
list of integers
\code{[1,~2]} and the second one to the integer
\code{3} ; the third and fourth arguments are not live, which is indicated
by the atom `-'.  There are two live local variables: \code{H}, which
is bound to the integer \code{1}, and \code{T}, which is bound to the
list of integers \code{[2]}. The goal path tells that this event
occurred in the \code{then} branch (\code{t}) of the second
conjunction (\code{c2}) of the first \code{switch} (\code{s1}) of
\code{partition/4}.

\end{enumerate}


\section{The Mercury queens program}
\label{queens-prog-appendix}

\begin{footnotesize}
\begin{verbatim}
:- module queens.

:- interface.

:- import_module io.

:- pred main(io__state, io__state).
:- mode main(di, uo) is cc_multi.

:- implementation.

:- import_module list, int.

main -->
    ( { data(Data), queen(Data, Out) } ->        
        io__write_string("A 5 queens solution is "), print_list(Out)
    ;
        io__write_string("No solution\n")
    ).

:- pred data(list(int)).
:- mode data(out) is det.

:- pred queen(list(int), list(int)).
:- mode queen(in, out) is nondet.

:- pred qperm(list(T), list(T)).
:- mode qperm(in, out) is nondet.

:- pred qdelete(T, list(T), list(T)).
:- mode qdelete(out, in, out) is nondet.

:- pred safe(list(int)).
:- mode safe(in) is semidet.

:- pred nodiag(int, int, list(int)).
:- mode nodiag(in, in, in) is semidet.

data([1,2,3,4,5]).

queen(Data, Out) :-
    qperm(Data, Out),
    safe(Out).

qperm([], []).
qperm([X|Y], K) :-
    qdelete(U, [X|Y], Z),
    K = [U|V],
    qperm(Z, V).

qdelete(A, [A|L], L).
qdelete(X, [A|Z], [A|R]) :-
    qdelete(X, Z, R).

safe([]).
safe([N|L]) :-
    nodiag(N, 1, L),
    safe(L).

nodiag(_, _, []).
nodiag(B, D, [N|L]) :-
    NmB is N - B,
    BmN is B - N,
    ( D = NmB ->
        fail
    ; D = BmN ->
        fail
    ;
        true
    ),
    D1 is D + 1,
    nodiag(B, D1, L).

:- pred print_list(list(int), io__state, io__state).
:- mode print_list(in, di, uo) is det.

print_list(Xs) -->
    ( { Xs = [] } ->
        io__write_string("[]\n")
    ;
        io__write_string("["),
        print_list_2(Xs),
        io__write_string("]\n")
    ).

:- pred print_list_2(list(int), io__state, io__state).
:- mode print_list_2(in, di, uo) is det.

print_list_2([]) --> [].
print_list_2([X|Xs]) --> 
    io__write_int(X),
    ( { Xs = [] } ->
        []
    ;
        io__write_string(", "),
        print_list_2(Xs)
    ).
\end{verbatim}
\end{footnotesize}


\label{lastpage}

\clearpage
\addcontentsline{toc}{section}{Index}
\flushbottom
\printindex

\end{document}